\documentclass{article}

\usepackage{PRIMEarxiv}

\usepackage[utf8]{inputenc} 
\usepackage[T1]{fontenc}    
\usepackage{hyperref}       
\usepackage{url}            
\usepackage{booktabs}       
\usepackage{amsfonts}       
\usepackage{nicefrac}       
\usepackage{microtype}      
\usepackage{lipsum}
\usepackage{graphicx}
\graphicspath{{media/}}     

\usepackage{etoolbox}
\usepackage{amsmath}
\usepackage{color}
\usepackage{hyperref}
\usepackage{tabularx}
\usepackage{makecell}
\usepackage[figuresleft]{rotating}
\usepackage{xcolor}
\usepackage{soul}

\title{A note on the error analysis of data-driven closure models for large eddy simulations of turbulence
}

\author{
  Dibyajyoti Chakraborty \\
  Pennsylvania State University\\ School of Information Sciences and Technology\\ University Park, PA, 16802. \\
  \texttt{d.chakraborty@psu.edu}
   \And
  Shivam Barwey \\
  Transportation and Power Systems Division\\
  240 Argonne National Laboratory\\
  Lemont, IL, 60439. \\
  \texttt{sbarwey@anl.gov}
  \AND
  Hong Zhang \\
  Mathematics and Computer Science Division\\ 240 Argonne National Laboratory\\
  Lemont, IL, 60439. \\
  \texttt{hongzhang@anl.gov} \\
  \And
  Romit Maulik \\
  Pennsylvania State University\\ School of Information Sciences and Technology\\ University Park, PA, 16802. \\
  \texttt{rmaulik@psu.edu} \\
}

\begin{document}
\include{vpack}
\maketitle

\begin{abstract}
In this work, we provide a mathematical formulation for error propagation in flow trajectory prediction using data-driven turbulence closure modeling. Under the assumption that the predicted state of a large eddy simulation prediction must be close to that of a subsampled direct numerical simulation, we retrieve an upper bound for the prediction error when utilizing a data-driven closure model. We also demonstrate that this error is significantly affected by the time step size and the Jacobian which play a role in amplifying the initial one-step error made by using the closure. Our analysis also shows that the error propagates exponentially with rollout time and the upper bound of the system Jacobian which is itself influenced by the Jacobian of the closure formulation. These findings could enable the development of new regularization techniques for ML models based on the identified error-bound terms, improving their robustness and reducing error propagation.
\end{abstract}




\section{Introduction}

Numerically simulating turbulent flow are challenging due to the multitude of temporal and spatial scales. Direct Numerical Simulation (DNS), which resolves all frequencies and wavenumbers in a turbulent flow, is often impractical for general use. In contrast, large eddy simulations (LES) resolves the lower wavenumbers (i.e., the larger structures in the flow-field) while modeling the affect of the unresolved scales. This is achieved by the use of LES closure models which are typically modeled as stresses and introduced as a forcing in the momentum equations through a divergence term. Lately, data-driven methods have increasingly been proposed for modeling LES closure with a particular emphasis on deep learning models \cite{sanderse2024scientific}.

A recent promising approach to data-driven closure modeling relies on the concept of differentiable physics \cite{list2022learned,shankar2023differentiable} where machine learning (ML) algorithms are embedded within discretized partial differential equation solvers and trained to minimize an \emph{a-posteriori} loss for the state trajectories. This is in contrast to other approaches which harvest training data for the subgrid tensor explicitly for training a model for the stress \cite{beck2019deep}. The latter approach necessitates assumptions for the nature of the filter and frequently requires ad-hoc treatments for stable deployment during \emph{a-posteriori} assessments. This is not observed with the former, since the influence of the numerical solver and its effective spatial and temporal discretization artifacts are built into the optimization formulation. 

The end result of a data-driven turbulence closure model is a hybrid physics and ML based technique to perform time-series predictions of turbulent flows. In such forecasting tasks, ML algorithms predict quantities in a closed-loop manner (frequently referred to as autoregressive forecasting). A hallmark of such a prediction mode is the propagation and amplification of small initial errors, leading to decreasing accuracy. Understanding and mitigating error propagation is crucial for developing robust models -- and in this study we perform a numerical analysis of a data-driven closure model that is trained using an \emph{a-posteriori} formulation and observe that errors in state predictions during autoregressive rollout can be upper-bounded. Furthermore, our analysis of this upper bound reveals an interaction between the Jacobian of the system as well as the error due to numerical time integration methods that may pave the way for improved regularization of data-driven models.

\section{Methods}
\subsection{Governing Equations}

Our governing equations are given by the incompressible Navier-Stokes equations,
\begin{equation}
\begin{aligned}
    \frac{\partial\mathbf{u}}{\partial t}+\nabla\cdot(&\mathbf{u}\otimes\mathbf{u})=\frac{1}{Re}\nabla^{2}\mathbf{u}-\frac{1}{\rho}\nabla p
    \\&\nabla\cdot\mathbf{u}=0,
\end{aligned}
\end{equation}
where $\mathbf{u}$ is the velocity vector, $p$ is the pressure, $\rho$ is the density, $Re$ is the Reynolds number, and we have ignored the presence of body forces. Typically, the Navier-Stokes equations are solved numerically on discretized domains to obtain an approximate solution given by $\mathbf{u^h}$ where $h$ represents the characteristic mesh size of the discretization.
The dimensionless parameter $Re$, representing the ratio of convective to diffusive forces in fluid flow, holds significant importance in characterizing various flow behaviors.
At high Reynolds numbers, inertial forces become dominant and can amplify small perturbations in the flow, generating energy-containing fluctuations or eddies across a wide range of spatial and temporal scales. When the high-wavenumber components of the solution field surpass the resolution of the numerical grid, direct numerical simulation (DNS) of the Navier-Stokes equations results in aliasing errors and inaccuracies in the solutions, typically manifesting energy pile-up near the grid cut-off wavenumbers. The only solution to this error that eschews closure modeling is the refinement of the grid to resolve the smallest physical length scales. Consequently, evaluating high Reynolds number flows with DNS can be cost-prohibitive or even infeasible for most practical applications.

In practical applications, there is a need to simulate high Reynolds number flows at resolutions that are computationally feasible, often considerably lower than those required by DNS. Large Eddy Simulation (LES) tackles this challenge by resolving the flow field on a coarse grid up to a defined cut-off length scale, represented by $\Delta$. In LES high-frequency contributions are filtered by the action of discretizing the governing equations on this coarse grid. This involves the assumption of a grid cut-off filter implicitly acting on the flow field, and resulting in a physical solutions without spurious artifacts stemming from insufficient resolution. In order to compute the filtered velocity field, the governing equations can be expressed using filtered variables, requiring the inclusion of an additional source term in the momentum equation. This term serves to accommodate the interactions between filtered and unfiltered scales within the flow field. The governing equations for LES are given by,
\begin{equation}{\frac{\partial{\overline{\mathbf{u}}}}{\partial t}}+\nabla\cdot({\overline{\mathbf{u}}}\otimes{\overline{\mathbf{u}}})={\frac{1}{Re}}\nabla^{2}{\overline{\mathbf{u}}}-{\frac{1}{\rho}}\nabla{\overline{p}}+\nabla\cdot{{\tau}}\end{equation}
where $\tau$ represents the impact of unresolved velocity components on the resolved field. This component, known as the subgrid-scale (SGS) stress, significantly influences the filtered velocity field. Neglecting this term implies the solution of an unresolved DNS which can result in numerical inaccuracies and potentially diverging flow fields. However, directly computation of $\tau$ is not feasible without access to the unresolved velocity components. Hence, it is necessary to prescribe a model to estimate $\tau$ based on the resolved field, thereby closing the problem. We reiterate that the concept of `filtering' a DNS field serves as a conceptual tool to describe the coarse-grained evolution within an LES. While the exact nature of this filter is often unknown, it can be approximated through prior knowledge of the numerical scheme employed (such as finite volume or spectral methods) usually for structured discretizations. However, in practice, the specifics of this filtering operation are generally obscure, posing challenges for data-driven modeling of $\tau$. Indeed, past work has shown that the choice of the filter significantly influences the calculation of $\tau$ \cite{piomelli1991subgrid}. This influences data-driven methods that seek to predict $\tau$ (i.e., \emph{a-priori} methods) where a poor assumption for the filter leads to poor results during \emph{a-posteriori} deployment \cite{Maulik2018}.

In response, there have been recent efforts to frame the learning of $\tau$ as PDE-constrained optimization problem \cite{shankar2023differentiable}. Given training data from DNS, the goal of such differentiable turbulence modeling methods is to learn a $\tau$ that produces an LES solution $\mathbf{\overline{u}}$ that minimizes the difference to DNS. We emphasize that this formulation \emph{does not require the generation of training data for $\tau$} but may be utilized to minimize errors with respect to quantities sampled (potentially sparsely) from the DNS simulations. In this note, we introduce an error analysis that demonstrates how this formulation can be leveraged to upper-bound the approximation error between the ground-truth solution and the LES solution as a function of the time step of a time integration scheme as well as the Jacobian of the governing equations and the ML model. We assume that the target trajectories to match are DNS state quantities sampled on an LES grid, independent of the assumption of a filter.

\section{Error Analysis}

The minimization objective of differentiable turbulence modeling is the prediction error at the $t^{th}$ time step which is given by some norm of the difference between the prediction using a data-driven closure and the ground truth. The error can be split into two terms signifying the error in approximating the training data (typically utilizing direct numerical simulations) using the best possible data-driven model, and the error in data-driven algorithm optimization for identifying the best model. We proceed by considering a simplified representation of the LES equations given by
\begin{equation}\label{PDE}
    \frac{d\mathbf{\overline{u}}}{dt} = F(\mathbf{\overline{u}}(\tau))
\end{equation}
where $\mathbf{\overline{u}}(\tau)$ is an LES solution obtained with a data-driven model predicting $\tau$. Note that this formulation allows for a more general interpretation of closure, i.e., we are not restricted to the explicit use of a stress term. The data-driven model is assumed to take flow field values, and additional physically relevant information, at the previous time step and predict a closure for the LES equations. The total prediction error is given by
\begin{equation}
\begin{aligned}
    \mathcal{E}_t = \|\mathbf{u}_t^h-\mathbf{\overline{u}}_t(\tau^k)\|= &\|\mathbf{u}_t^h-\mathbf{\overline{u}}_t(\tau^*)+\mathbf{\overline{u}}_t(\tau^*)-\mathbf{\overline{u}}_t(\tau^k)\|\\
    &\leq \underbrace{\| \mathbf{u}_t^h-\mathbf{\overline{u}}_t(\tau^*)\|}_\text{$\beta_t$}+\underbrace{\|\mathbf{\overline{u}}_t(\tau^*)-\mathbf{\overline{u}}_t(\tau^k)\|}_\text{$\gamma_t$},
\end{aligned}
\end{equation}
where $\mathbf{u}_t^h$ is a state constructed by sampling DNS on an LES grid. When grid points from DNS are collocated with those of LES - no assumption of filtering is made for the subsequent error analysis. When collocation is not possible and interpolation is necessary, an effective filtering operation on the state is implied. However, we note that this filtering operation is not used to recover quadratic subgrid interactions and merely serves as a means to generate training data for the ultimate solver and data-driven closure combination (a hallmark of \emph{a-posteriori} data-driven closure modeling). Next, $\mathbf{\overline{u}}_t(\tau^*)$ is an LES solution with the best possible data-driven turbulence model at the $t^{th}$ time step, and $\mathbf{\overline{u}}_t(\tau^k)$ is the same model at the k$^{th}$ optimization step. We define $\beta_t$ as the approximation error for the best possible ML closure model given a coarse discretization, machine learning algorithm and hyperparameters, and $\gamma_t$ is the optimization error which includes the error due to not identifying the best possible trainable parameters of the ML model. We emphasize that $\beta_t$ represents the error between the ground truth state trajectories and the LES solution on the coarse-grid. Furthermore, $\gamma_t$ is introduced since most optimized data-driven models are not expected to reach the global minima. 

\subsection{Modeling Error}
The error in approximating DNS on the LES grid with the optimal machine learning closure is given by
\begin{equation}
\begin{aligned}
    \beta_t = \| \mathbf{u}_t^h-\mathbf{\overline{u}}_t(\tau^*)\|
\end{aligned}
\end{equation}
Assuming the use of an explicit Euler based integration of the discretized LES equations, the local truncation error at the t$^{th}$ time step may be represented as
\begin{equation}\label{bound 1}
    E_t^* = \mathbf{\overline{u}}_t(\tau^*) - \mathbf{\overline{u}}_{t-1}(\tau^*) - \Delta t F(\mathbf{\overline{u}}_{t-1}(\tau^*))
\end{equation}
It is well known that $E_t^*$ is bounded as follows \cite{davis2007methods}
\begin{equation}
    \| E_t^* \| \leq \frac{M^*}{2}(\Delta t)^2 
\end{equation}
for some constant $M^*$. A similar result may be obtained for the evolution of the ground truth data on the LES grid, i.e $u^{h}$ satisfies Equation \ref{PDE} for some ideal (but unknown) stress $\tau^h$.
\begin{equation}\label{bound*}
    E_t^h = \mathbf{u}_t^h - \mathbf{u}_{t-1}^h - \Delta t F(\mathbf{u}_{t-1}^h(\tau^h)) \leq \frac{M^h}{2}(\Delta t)^2, 
\end{equation}
where we note an assumption that this evolution also utilizes an explicit Euler integration with an identical timestep. Therefore, by using equation \ref{bound 1} and \ref{bound*} ,
\begin{equation}\label{e2}
\begin{aligned}
    \beta_t &\leq \| \mathbf{u}_{t-1}^h + \Delta tF(\mathbf{u}_{t-1}^h (\tau^h)) + E_t^h - \mathbf{\overline{u}}_{t-1}(\tau^*) -  \Delta t F(\mathbf{\overline{u}}_{t-1}(\tau^*)) - E_t^*\|\\
\end{aligned}
\end{equation}
The mean value theorem states that,
\begin{equation}
\min_{a\leq x\leq b}f'(x)\leq\frac{f(b)-f(a)}{b-a}\leq\max_{a\leq x\leq b}f'(x)
\end{equation}
So,
\begin{equation}\label{MVT}
    \|F(\mathbf{u}_{t-1}^h(\tau^h)) -  F(\mathbf{\overline{u}}_{t-1}(\tau^*))\| \leq \beta_{t-1} \max_{\mathbf{u}_{t-1}^h (\tau^h)\leq \mathbf{\overline{u}}\leq \mathbf{\overline{u}}_{t-1}(\tau^*)}\|\frac{dF}{d\mathbf{\overline{u}}}\|
\end{equation}
Using equation \ref{e2} and \ref{MVT}, we get 
\begin{equation}\label{e_t}
    \beta_t \leq (1+ R_\beta \Delta t)\beta_{t-1} + C_\beta(\Delta t)^2
\end{equation}
where $R_\beta$ is the upper bound of the Jacobian norm $\|\frac{dF}{d\mathbf{\overline{u}}}\|$ for state vectors in the domain $[\mathbf{u}_t^h (\tau^h), \mathbf{\overline{u}}_t(\tau^*)]$ for all $t$, and $C_\beta$ is $\|\frac{(M^h-M^*)}{2}\|$. We note that the computation of the Jacobian using $\mathbf{u}_h$ can be performed on appropriate locations of the DNS grid and projected onto the LES grid for the computation of $R_\beta$ (in a manner similar to training data generation). Applying Equation \ref{e_t} recursively we get
\begin{equation}
    \beta_t \leq c^t\beta_0 + (1+c+c^2+c^3...c^{t-1})C_\beta(\Delta t)^2
\end{equation}
where $c$ is $(1+R_\beta\Delta t)$ and $\beta_0$ is the error in first time step. If we interpret the training of an ML model for $\tau^*$ (embedded in our LES) as an optimization that propagates gradients through the discretized system of equations, we can leverage the fact that our optimization is devised to drive down the prediction error (of the state) for a one timestep prediction to a small constant error $\epsilon$. Intuitively, we expect that this small error to be compounded with autoregressive rollouts as we deploy our closure for greater prediction horizons. Considering a prediction horizon given by $T=t\Delta t$, 
\begin{equation}
    (1+R_\beta\Delta t)^t = \left (1+\frac{R_\beta T}{t}\right)^t\leq e^{R_\beta T}
\end{equation}
and
\begin{equation}
\begin{aligned}
    1+c+c^2 \dots c^{t-1}&=\frac{(1+R_\beta\Delta t)^t-1}{R_\beta\Delta t}\\
    &\leq t\frac{e^{R_\beta T}-1}{R_\beta T}\\
\end{aligned}
\end{equation}
Which finally gives us,
\begin{equation}
\begin{aligned}
    \beta_t &\leq e^{R_\beta T}\epsilon + \frac{e^{R_\beta T}-1}{R_\beta}C_\beta \, \Delta t\\
\end{aligned}
\end{equation}

\subsection{Optimization Error}
The optimization error encapsulates inadequate training of our differentiable physics model, such as when optimization is truncated before a global minima is reached. It is defined as
\begin{equation}
    \gamma_t = \|\mathbf{\overline{u}}_t(\tau^*)-\mathbf{\overline{u}}_t(\tau^k)\|
\end{equation}
Expanding this in a manner similar to the modeling error we get,
\begin{equation}
    \gamma_t \leq (1+R_\gamma\Delta t)\gamma_{t-1} + C_\gamma(\Delta t)^2
\end{equation}
where $R_\gamma$ is assumed as the upper bound of the Jacobian $\|\frac{dF}{d\mathbf{\overline{u}}}\|$ in the domain $[\mathbf{\overline{u}}_t(\tau^*), \mathbf{\overline{u}}_t(\tau^k)]$ for all $t$, and $C_\gamma$ is $\|\frac{(M^*-M^k)}{2}\|$. Using a recursive relation similar to modeling error we get,
\begin{equation}
\begin{aligned}
    \gamma_t \leq e^{R_\gamma T}\gamma_0+ \frac{e^{R_\gamma T}-1}{R_\gamma}C_\gamma \, \Delta t
\end{aligned}
\end{equation}
where $\gamma_0$ is the optimization error for the first time step which is bounded\cite{khara2024neural} by 
\begin{equation}\label{opt_bound}
    \gamma_0 \leq\left(1-\frac2{\kappa+1}\right)^{2N}\|\theta_0-\theta^*\|^2
\end{equation}
where $\tau \equiv \tau(\mathbf{\overline{u}},\theta)$ is the neural network that predicts the closure stresses given trainable parameters $\theta$, $N$ is the number of the epochs in the optimization, $\kappa = \mu/\beta$ assuming a $\mu-$strong convex function that is $\beta$ smooth. 
\subsection{Total Error}
Finally we note that the total approximation error is bounded by,
\begin{equation}
    \mathcal{E}_t \leq e^{R_\beta T}\epsilon + \frac{e^{R_\beta T}-1}{R_\beta}C_\beta \, \Delta t + e^{R_\gamma T}\gamma_0+ \frac{e^{R_\gamma T}-1}{R_\gamma}C_\gamma \, \Delta t
\end{equation}
Considering $R$ as the global upper bound of the Jacobian norm $\|\frac{dF}{d\mathbf{\overline{u}}}\|$, it can be further simplified to
\begin{equation}
    \mathcal{E}_t \leq e^{R T}(\epsilon+ \gamma_0) + \frac{e^{R T}-1}{R} (C_\beta+C_\gamma) \Delta t
\end{equation}
This shows that the upper bound of the error at a time $t$ grows exponentially with the upper bound of the Jacobian of the solver and the total simulation time $T=t\Delta t$. It also grows linearly with the corresponding error terms($\epsilon$ and $\gamma_0$) at the first time step.
\section{Analysis}


Our error approximation analysis reveals the following information about data-driven closure models for large eddy simulations. The analysis explicitly quantifies the recursive nature of error growth during autoregressive (i.e., rollout) deployments in a-posteriori assessments. This has been observed, empirically, in several applications but is rigorously quantified through our derivations. Furthermore, we observe that error growth during deployment is explicitly connected to the Jacobian $\frac{dF}{d\mathbf{\overline{u}}}$ that is also influenced by $\frac{\partial \tau}{\partial \mathbf{\overline{u}}}$. This may be useful for regularization of neural network models during training. Another important finding is that magnitude of the step size, a consequence of temporal discretization, also contributes to error growth. We also note that when deep learning based models for $\tau$ are used, it may be difficult to ascertain the effect of the optimization error due to non-convexity. This may motivate the use of strictly convex function approximation for learning such closures \cite{amos2017input}. For extending the present analysis to increasingly realistic cases we identify the following opportunities for improvement. Firstly, our analysis is introduced for a relatively simple explicit time integration scheme with a fixed step size shared across DNS and LES. This can be improved in subsequent studies. Additionally, adaptive step size integrators that are potentially higher-order or implicit must also be studied for their effect of autoregressive error. We also note that for most applications, DNS is not available, and it is impossible to determine the norm of the Jacobian of the DNS data on an LES grid. Therefore, the upper bound $R_\beta$ may well be approximate in practice. However, controlling the Jacobian of the neural network can be used indirectly for reducing autoregressive error growth.

\section{Conclusion}
In this work we derive a mathematical formulation for the error propagation in prediction of flow trajectories for LES using data-driven closure models. We find out that the error depends on the initial prediction error at the first step, the time step for integration, and a combination of the Jacobian of the solver or the data-driven ML model depending on the formulation. We also show that the error propagates exponentially with autoregressive rollouts and hypothesize potential regularization strategies for enhancing the accuracy of the data-driven LES. This work represents a first example of a rigorous characterization of the interaction between LES and data-driven turbulence models in terms of a-posteriori error.

\subsection*{Acknowledgements}

This material is based upon work supported by the U.S. Department of Energy (DOE), Office of Science, Office of Advanced Scientific Computing Research, under Contract No. DE-AC02–06CH11357. 
RM and DC acknowledge support from DOE ASCR award ``Inertial neural surrogates for stable dynamical prediction" and DOE FES award ``DeepFusion Accelerator for Fusion Energy Sciences in
Disruption Mitigation". SB acknowledges support from the AETS Fellowship at Argonne National Laboratory. HZ acknowledges support from the Scientific Discovery through Advanced Computing (SciDAC) program through the FASTMath Institute under contract DE-AC02-06CH11357 at Argonne National Laboratory. This paper describes objective technical results and analysis. Any subjective views or opinions that might be expressed in the paper do not necessarily represent the views of the U.S. DOE or the United States Government.

\bibliographystyle{unsrt}  
\bibliography{bib}

\end{document}